\documentclass{nature1}
\usepackage{graphicx}
\title{Type Ia Supernovae as Stellar Endpoints\\ and Cosmological Tools}

\author{D. Andrew Howell$^{1,2}$}

\usepackage[super]{natbib}

\begin{document}

\maketitle

\begin{affiliations}
\item Las Cumbres Observatory Global Telescope Network,
 6740 Cortona Dr., Suite 102, Goleta, CA 93117
\item Department of Physics, University of California,
 Santa Barbara, Broida Hall, Mail Code 9530, Santa Barbara, CA 93106-9530
\end{affiliations}

\newcommand\ion[2]{#1$\;${\small\rmfamily\@Roman{#2}}\relax}%
\newcommand\nodata{ ~$\cdots$~ }%
\newcommand\arcdeg{\mbox{$^\circ$}}%
\newcommand\degr{\arcdeg}%
\newcommand{\etal}{et~al.\,}
\newcommand{\Msun}{${\rm M}_\odot$}          
\newcommand{\drp}{$\Delta m_{15}(B)$} 
\newcommand{\Ni}{$^{56}\rm Ni$} 
\newcommand{\Co}{$^{56}\rm Co$} 
\newcommand{\Fe}{$^{56}\rm Fe$} 
\newcommand{\kps}{km s$^{-1}$}
\newcommand{\kms}{km s$^{-1}$}
\newcommand{\MCh}{$M_{\rm Ch}$}
\newcommand{\hof}{H\"oflich}
\newcommand{\eg}{e.\,g.\,}
\newcommand{\ie}{i.\,e.\,}
\newcommand{\OI}{\ion{O}{1}}
\newcommand{\SiII}{\ion{Si}{2}}
\newcommand{\Si}{\ion{Si}{2}}
\newcommand{\Simain}{\ion{Si}{2} 6150\AA}
\newcommand{\Sifour}{\ion{Si}{2} 4000\AA}
\newcommand{\SII}{\ion{S}{2}}
\newcommand{\CaII}{\ion{Ca}{2}}
\newcommand{\Camain}{\ion{Ca}{2} 3800\AA}
\newcommand{\Ca}{\ion{Ca}{2}}
\newcommand{\TiII}{\ion{Ti}{2}}
\newcommand{\MgII}{\ion{Mg}{2}}
\newcommand{\iab}{i\arcmin (AB)}
\newcommand{\gt}{$>$}
\newcommand{\us}{$u$*}
\newcommand{\gp}{$g$\arcmin}
\newcommand{\rp}{$r$\arcmin}
\newcommand{\ip}{$i$\arcmin}
\newcommand{\zp}{$z$\arcmin}
\newcommand\arcmin{\mbox{$^\prime$}}%
\newcommand\arcsec{\mbox{$^{\prime\prime}$}}%
\newcommand\araa{Ann. Rev. Astron. Astrophys.}%
\newcommand\apj{Astrophys. J.}%
\newcommand\aj{Astron. J.}%
\newcommand\apjl{Astrophys. J. Lett.}%
\newcommand\nar{New Astron. Rev.}%
\newcommand\apjs{Astrophys. J. Suppl.}%
\newcommand\apss{Astrophys. Sp. Sci..}%
\newcommand\aap{Astron. \& Astrophys.}%
\newcommand\mnras{MNRAS}%
\newcommand\nat{Nature}%
\newcommand\iaucirc{IAU~Circ.}%
\newcommand\gca{Geochim.~Cosmochim.~Acta}%
\newcommand\aaps{Astron. \& Astrophys. Suppl. Ser.}%
\newcommand\pasp{PASP}
\newcommand\pasj{PASJ}

\begin{abstract}
  Empirically, Type Ia supernovae are the most useful, precise, and
  mature tools for determining astronomical distances.  Acting as
  calibrated candles they revealed the presence of dark energy and are
  being used to measure its properties.  However, the nature of
  the Type Ia explosion, and the progenitors involved, have remained
  elusive, even after seven decades of research.  But now new large
  surveys are bringing about a paradigm shift --- we can finally
  compare samples of hundreds of supernovae to isolate critical
  variables.  As a result of this, and advances in modeling,
  breakthroughs in understanding all aspects of these supernovae are finally
  starting to happen.
\end{abstract}

``Guest stars'' (who could have imagined they were distant stellar
explosions) have been surprising humans for at least 950 years, but
probably far longer.  They amazed and confounded the likes of Tycho,
Kepler, and Galileo, to name a few.  But it was not until the
separation of these events into novae and supernovae by Baade and
Zwicky that progress understanding them began in
earnest\citep{1934PhRv...46...76B}.  This process of splitting a
diverse group into related subsamples to yield insights into their
origin would be repeated again and again over the years, first by
Minkowski when he separated supernovae of Type I (no hydrogen in their
spectra) from Type II (have hydrogen)\citep{1941PASP...53..224M}, and
then by Elias et al. when they determined that Type Ia supernovae (SNe
Ia) were distinct\citep{1985ApJ...296..379E}.  We now define Type Ia
supernovae as those without hydrogen or helium in their spectra, but
with strong SiII, which has observed absorption lines at 6150, 5800
and 4000\AA\ \citep{1997ARA&A..35..309F}.  The progenitor of a Type Ia
supernova has never been seen, but the consensus is that they are the
result of the thermonuclear explosion of a degenerate carbon-oxygen
white dwarf (WD) star\citep{1960ApJ...132..565H} gaining mass in a binary
system\citep{1973ApJ...186.1007W}, in contrast to the other types of
supernovae, which are thought to result from the collapse of the core
of a massive star.  Much about them remains unknown --- the process of
categorizing and trying to understand them continues.

As distance indicators, SNe Ia are capable of better than 6\% precision in
distance after empirical corrections are made for their lightcurve
shape and color.  Just over a decade ago, they were used to deliver
perhaps the most shocking scientific discovery in the last
half-century: the expansion of the universe is not slowing, but is
instead accelerating\citep{1998AJ....116.1009R,1999ApJ...517..565P}.
The ``dark energy'' required to power this acceleration is prodigious,
and indeed must be the dominant constituent of the universe by a
factor of 17 over normal matter, and 3 over dark matter (which,
despite the name, is probably unrelated).  But because dark energy is so dilute
(one cubic meter contains the energy equivalent to a few atoms of
hydrogen), its effects can only be seen on vast scales.  We must
compare supernovae across $\sim 10$ Gyr of cosmic time, during which
stars and galaxies have undergone substantial evolution.  Therefore it
is critical to understand supernovae if they are to be relied on as
the primary distance probe in astrophysics.

We bein by reviewing SN Ia lightcurves and spectra (\S\ref{lcspecsec})
and their use in cosmology (\S\ref{cosmologysec}).  This has been
refined to the point that SNe Ia are now sensitive to subtle,
systematic effects (\S\ref{syssec}).  
However, many of these are
being addressed by comparing large subsamples of hundreds of SNe Ia,
possible for the first time only in the last few years.

New multidimensional simulations are revealing
once-elusive details of the explosion, such as how asymmetry
contributes to the diversity seen in SNe Ia (\S\ref{theorysec}).
Advances in understanding progenitors have been proceeding apace, led
by new observations.  The properties of SNe Ia
are being correlated with the stellar populations in their host
galaxies, yielding real progenitor constraints (\S\ref{progensec}).
Dramatic new discoveries, like supernovae that appear to require a
progenitor above the Chandrasekhar mass, are challenging pre-existing
theories.  Evidence has been mounting that the long-favored scenario
of accretion from a main sequence or red giant companion star cannot
account for all SNe Ia.  Instead, the white
dwarf merger and even the sub-Chandrasekhar mass explosion models have
been experiencing a revival.

Finally, we conclude that empirical SN Ia cosmology remains solid, but
is being continually refined by taking into account ever-more-subtle
effects and new knowledge.  But our understanding the the progenitors
of SNe Ia and the explosion process may be undergoing an evolution, as
the data from vast new surveys with thousands of supernovae are
becoming available (\S\ref{finalsec}).

\section{Lightcurves and spectra}\label{lcspecsec}
In a SN Ia, a wave of thermonuclear fusion rips through a degenerate
white dwarf star, synthesizing iron-peak elements (Ni, Co, Fe) in the
dense inner regions, intermediate mass elements (Si, S, Ca, Mg, O)
where burning is incomplete, and sometimes leaving unburned material
(C,O) near the outer layers\citep{2000ARA&A..38..191H}.  Though the
explosion provides the kinetic energy of the SN Ia, and unbinds the
white dwarf, this is not what we see as the supernova.  The lightcurve
is powered by the radioactive decay of $^{56}$Ni (half-life 6.1 days)
to $^{56}$Co, and ultimately to $^{56}$Fe (half-life 77 days).  Gamma
rays produced in the decays are thermalized, and at peak light $\sim
85\%$ of the light output of the SN Ia is in the optical, peaking at
4000\AA , with the remainder mostly radiated in the near-ultraviolet
(NUV) and near-infrared (NIR)\citep{2009ApJ...691..661H}.

\begin{figure}
\includegraphics[width=7in]{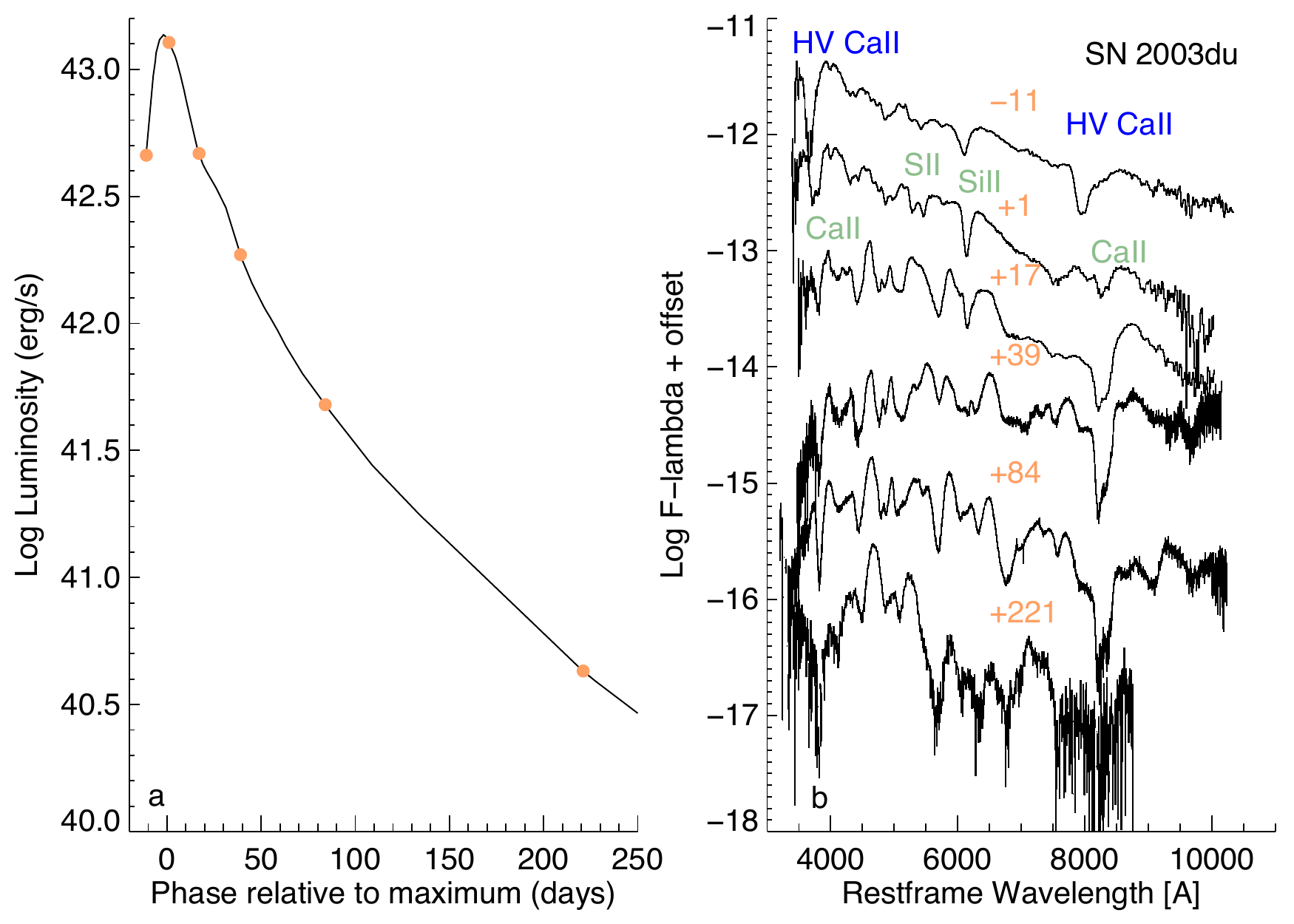} 
\caption{\textsf{ {\bf \textsf{UVOIR lightcurve and selected spectra from the normal Type Ia supernova SN 2003du.}}  (a) Quasi-bolometric (UV-optical-infrared)
  lightcurve for SN 2003du\citep{2007A&A...469..645S}.  Orange points
  mark the dates of spectra shown in panel b. (b) Spectra of SN
  2003du\citep{2007A&A...469..645S}, with phases relative to maximum
  light listed in orange.  Characteristic intermediate mass element
  features seen in maximum light spectra of SNe Ia are noted in green:
  CaII, SII, and SiII.  Before maximum light, when we are seeing the
  outermost layers of the ejecta, CaII is seen at high velocity, a
  common feature in SNe Ia.  SN ejecta thin with time, and we see to
  deeper layers.  As SNe leave the photospheric phase and enter the
  nebular phase, emission features start to dominate.}}
\label{fluxspec}
\end{figure}

Figure~\ref{fluxspec} shows the quasi-bolometric (UV-optical-IR)
lightcurve evolution of a typical SN Ia, SN
2003du\citep{2007A&A...469..645S}, along with representative spectra.
SNe Ia rise to maximum light over $\sim 13-23$ days in the $B$-band,
with a typical rise time of about 17.5
days\citep{2010ApJ...712..350H}.  Initially, the time for photons to
diffuse out of the dense ejecta is high, so that the rate of
deposition of energy by radioactive decay exceeds the energy radiated
by the SN.  Thus, at early times, spectra show only absorption lines
probing the outer layers of the SN.  As the SN expands and thins, the
light-emitting region (the photosphere) recedes (in mass or velocity
coordinates), and spectra start to probe deeper layers.  Eventually
the ejecta reach a point where the rate of energy deposition by
radioactive decay is equal to the radiated luminosity of the SN, and
the SN starts to decline in luminosity\citep{1982ApJ...253..785A}.
Around or after maximum light, spectra show P-Cygni profiles --
emission at the rest wavelength (initially weak), and blueshifted
absorption.  Weeks after maximum light, the spectra start becoming
dominated by scattering from permitted lines (at least in the
blue)\citep{2008PASP..120..135B} and ultimately become dominated by
emission features as the ejecta start to become optically thin, and
the SN makes the transition from the photospheric to the nebular
phase.  At late times (hundreds of days), gamma rays can freely
escape, but positrons may be trapped\citep{2006AJ....132.2024L}, and
the lightcurve slope may match the slope expected from the decay of
$^{56}$Co, or can be steeper than this in the case of incomplete
trapping\citep{2009A&A...505..265L}.  Late time spectra are dominated
by emission lines from iron-peak elements, synthesized in the deepest
regions where the white dwarf was densest.

\section{Cosmology}\label{cosmologysec}
\begin{figure}
\includegraphics[width=7in]{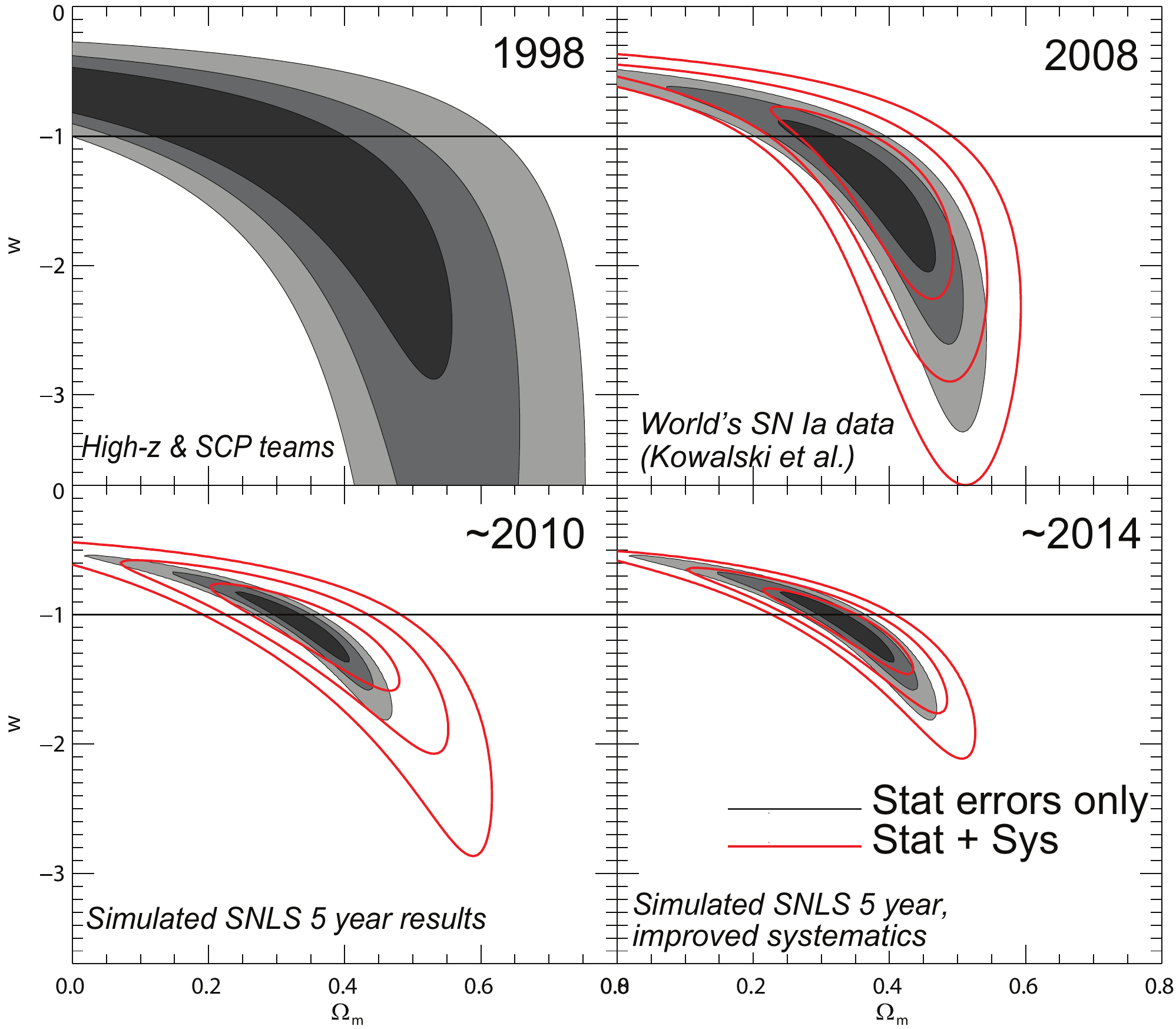} 
\caption{\textsf{ {\bf\textsf{Improvements in dark energy equation of state
    constraints from Type Ia supernovae over time.}}  The top two
  panels show the remarkable improvements in $w$ made using SNe Ia
  over the past decade (assuming a flat universe), the bottom two show
  the importance of improving systematics in the early years of the
  next decade\citep{2009arXiv0903.1086H}.  (a) $\Omega_M - w$
  statistical-only constraints circa 1998\citep{1998ApJ...493L..53G}.
  (b) By 2008, Kowalski et al., combining many data sets, showed
  that systematic uncertainties are
  significant\citep{2008ApJ...686..749K}. (c) Expected
  constraints for the year 5 results of SNLS, assuming additional
  low-z SNe, and double the number of $z > 1$ SNe from HST, assuming
  there is no improvement in systematic uncertainties from the 3rd
  year result.  (d) Assumes the low-z data are on the
  Sloan photometric system, and a factor of two improvement in
  measurements of fundamental flux standards.  Judged by the area of
  the inner 68.3\% contour, the improvement from the 1998 results is a
  factor of 3, 5, and 10, including systematics.}}
\label{wplot}
\vskip 0.05 in
\end{figure}

Dark energy is often characterized by its ratio of pressure to
density, i.e., its equation of state, $w=P/\rho c^2$
(Fig.~\ref{wplot}).  If $a$ is an arbitrary length scale in the
universe, then the density of some component of the universe evolves
as $\rho \propto a^{-3(1+w)}.$ Normal matter has $w=0$; it dilutes
with volume as space expands.  If dark energy has an equation of state
$w=-1$, corresponding to Einstein's cosmological constant, $\Lambda$,
then it has the strange property that the energy density does not
dilute as the universe expands; it must be a property of the vacuum.
Other possibilities exist: if $w<-1$ the dark energy density is
ever-growing, and will ultimately result in a ``big rip,'' destroying
galaxies and even subatomic particles.  However, $w$ can also be a
scalar field with $w>-1$ (generally known as quintessence), and the
value of $w$ can evolve over time.

Measurements of $\langle w \rangle$ or $w(z)$ are obtained by building
a map of the history of the expansion of the universe using SNe Ia as
standardized candles.  SNe Ia can show a factor of 10 or more
difference in peak luminosity, but the luminosity is correlated with
the time it takes the supernova to rise and fall in
brightness\citep{1993ApJ...413L.105P}.  Therefore, the width of the
lightcurve is measured and used to correct the peak luminosity.
Common parametrizations include the ``stretch
factor\citep{1999ApJ...517..565P}'', $s$, proportional to lightcurve
width, or \drp , the number of magnitudes the $B$-band lightcurve
falls in 15 days after peak brightness, inversely proportional to
lightcurve width\citep{1993ApJ...413L.105P}.  A correction must also
be made for color, since redder SNe are dimmer, both intrinsically,
and due to dust\citep{1996ApJ...473...88R,2007ApJ...664L..13C}.
Various techniques are used to determine these parameters by fitting
lightcurve models to the data, but leading fitters include
MLCS2k2\citep{2007ApJ...659..122J}, SALT2\citep{2007A&A...466...11G},
and SiFTO\citep{2008ApJ...681..482C}.

To determine cosmological parameters, an observed Hubble diagram
(distance versus redshift) is constructed, and cosmological parameters
are varied in a model that is fitted to the data.  If we express
distances as magnitudes, as a distance estimator, we can
use\citep{2006A&A...447...31A}:
\begin{equation}
\mu_B=m^*_B-M+\alpha (s-1)-\beta c,
\end{equation}
where $m^*_B$ is the peak magnitude of the SN in the $B$-band (the
blue filter where the SN is brightest), $s$ is the lightcurve stretch,
$c$ is the color of the SN (a linear combination of $U-B$ and $B-V$,
relative to some reference
color\citep{2006A&A...447...31A,2008ApJ...681..482C}), $\alpha$ is the
slope of the stretch-luminosity relation, $\beta $ is the slope of the
color-luminosity relation, and $M$ is a measure of the absolute
magnitude of the SN combined with the Hubble constant.  Note that
since relative magnitudes are used, neither the absolute magnitude of
the SN, nor the value of the Hubble constant must be known.  $m^*_B$,
$s$, and $c$ are measured from a fit to each SN lightcurve, while $M$,
$\alpha $, and $\beta$ are constants determined from the overall
cosmological fit.  


\section{Systematic uncertainties}\label{syssec}
Cosmological studies using SNe Ia, have reached a point where over
most redshift ranges systematic errors (those that affect many
measurements simultaneously in a correlated way) dominate statistical
errors (i.e. those that are reduced by
$\sqrt{N}$)\citep{2009ApJS..185...32K, 2009ApJ...700.1097H,
  2010ApJ...716..712A} --- see Fig.~\ref{wplot}.  
Many
systematic uncertainties can be lowered with new methods, improved
statistics, by comparing subsamples of supernovae, or breakthroughs in
understanding.  The primary systematic uncertainties
affecting SNe Ia are survey-dependent, but
there is general consensus that the largest are: calibration to the
historic Landolt photometric system, treatment of the ultraviolet,
reddening due to dust, the differences supernovae show with respect to
environment, and the possible evolution of SNe with redshift.  See
Table 1 for a summary of recent SN Ia constraints on $\langle w
\rangle$ and dominant systematic uncertainties.

\subsection{Calibration}
As supernovae are observed at high redshift, the region of the
spectrum seen through a given broadband filter changes.  A correction
must be made for this, the k-correction, which requires
knowledge of the spectral energy distribution of an average
supernova\citep{2007ApJ...663.1187H}, and precise knowledge of the
filter transmission curves and calibration system used for each
supernova.  Unfortunately, many historical low-redshift supernovae
were transformed to the Landolt photometric standard system, which is
poorly understood and no longer reproducible, as the filter and
instrument transmission curves used to establish it are not known to 
modern precision and no longer exist.  The fact that the low-z SN sample
is on the Landolt system has forced many high redshift surveys to
transform to it, incurring a systematic uncertainty in zero points and
color terms\citep{2006A&A...447...31A}.  These systematic
uncertainties will be reduced once a large new low-redshift sample of
SNe is assembled, calibrated onto a new system (e.g. Sloan).  

\subsection{The Ultraviolet}\label{uv}
At redshifts above $z\sim 0.2$, the restframe $U$-band and
ultraviolet is redshifted into the optical region of the spectrum.
Unfortunately, there is a great deal of scatter in restframe $U$
observations.  Some surveys rely heavily on the restframe $U$, so uncertainties from this region can translate into dominant
systematic uncertainties\citep{2009ApJS..185...32K}.  In fact, differences
seen between the distances determined by different lightcurve fitters,
e.g MLCS2k2\citep{2007ApJ...659..122J},
SALT2\citep{2007A&A...466...11G}, and SiFTO\citep{2008ApJ...681..482C}
boil down largely to the way they treat the restframe $U$-band.

There are two possibilities for $U$-band uncertainty: (1) SNe
intrinsically show more variation in $U$\citep{2008ApJ...674...51E},
and/or (2) the $U$-band is poorly calibrated.  The Supernova Legacy Survey (SNLS) has shown that
when longer-wavelength high redshift observations from a
well-calibrated single survey are k-corrected to restframe $U$-band,
the dispersion is more than 3 times lower than for low-redshift data
where the $U$-band is observed directly\citep{2006A&A...447...31A}.
Therefore, discrepancies are most likely due to problematic low
redshift observations where many factors conspire to make $U$
observations difficult, including atmospheric variation, extinction,
nonstandard filters, and poor calibrators\citep{2006AJ....131..527J}.
This systematic can be dealt with by building a new, better-calibrated
low redshift sample, and possibly by avoiding the restframe $U$-band.

\subsection{Reddening}\label{reddening}
Lightcurve fitters must correct for the fact that redder supernovae
are dimmer.  This is due to a combination of an intrinsic
color-luminosity relation (faint supernovae are intrinsically
red\citep{1996ApJ...473...88R}), and reddening due to dust.  While
there is consensus that the two effects ought to be corrected for
independently, there is disagreement over whether this can be
practically achieved given current data limitations.

\begin{figure}
\includegraphics[width=5in]{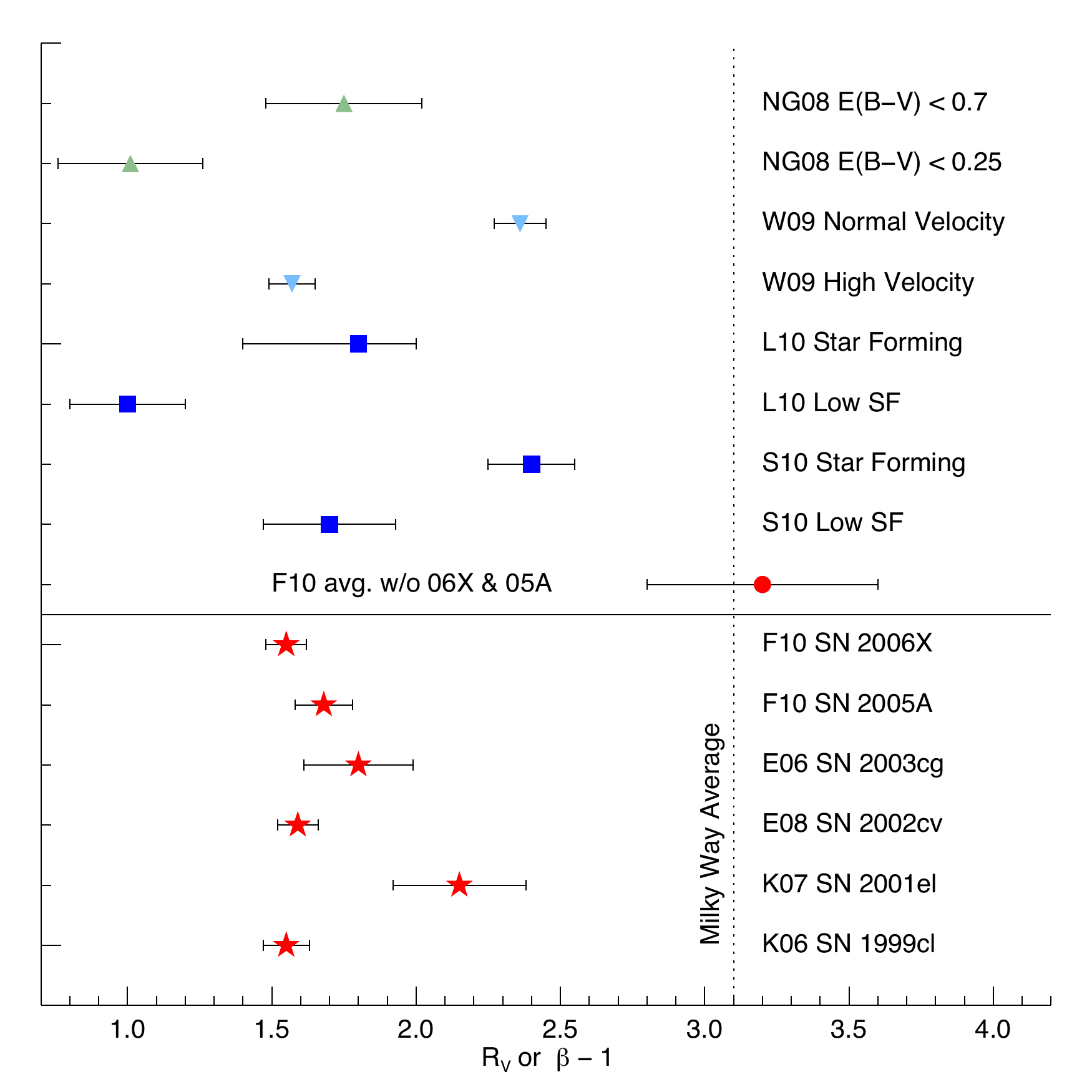} 
\caption{\textsf{{\bf \textsf{Determination of the reddening law along the line of
    site to SNe Ia.}}  In most cases $R_V=A_V/E(B-V)$ is shown, but in
  the S10 data, $\beta$-1 is shown, equivalent to $R_V$ if
  the colors of supernovae after lightcurve shape correction are due
  to dust reddening.  Different symbols/colors correspond to different
  techniques.  The most reliable is using $u$ to IR data to determine
  the reddening law (red stars), but this can only be done for a few
  individual SNe Ia.  Data is from K06\citep{2006AJ....131.1639K},
  K07\citep{2007AJ....133...58K}, E06\citep{2006MNRAS.369.1880E},
  E08\citep{2008MNRAS.384..107E}, and F10\citep{2010AJ....139..120F}.
  The average from the Carnegie Supernova Program (F10) IR data,
  excluding their most reddened events, is shown as a red circle, the
  only determination consistent with Milky Way dust (dotted line).
  However, NG08\citep{2008A&A...487...19N} (green triangles) come to
  the opposite conclusion (albeit using only UBVRI data): when they
  restrict themselves to the least reddened supernovae, they find a
  lower value of $R_V$.  Blue squares show reddening behavior
  determined by solving for the reddening law for ensembles of SNe
  that would minimize their residual on the Hubble diagram.  Results
  split by star forming galaxies and those with low star formation are
  shown from the SDSS-II SN Survey (L10\citep{2010ApJ...722..566L})
  and the SNLS (S10\citep{2010MNRAS.406..782S}).  There are some
  indications that solving for reddening by minimizing Hubble
  residuals is incompatible with IR
  determinations\citep{2010AJ....139..120F}.  Finally,
  W09\citep{2009ApJ...699L.139W} find that supernovae with high
  velocity features in their spectra have a lower $R_V$, perhaps an
  indication that there is a component of reddening that is intrinsic
  to the SN. }}
\label{redfig}
\end{figure}

MLCS2k2 attempts to separate intrinsic and dust reddening and make
different corrections for each\citep{2007ApJ...659..122J}.  
SALT2 and SiFTO work under the assumption that it is difficult or
impossible to separate the effects of dust and intrinsic reddening, so
they make an empirical reddening correction by solving for the slope
of the color-luminosity relation, $\beta $.  If this reddening were
only due to dust, then $\beta=R_B=R_V+1.$ For Milky Way dust,
$R_B=4.1$.  However, as shown in
Fig.~\ref{redfig}, shallower slopes are found for
$\beta$.  This may result from the
conflation of dust and intrinsic SN reddening, and would indicate that
the intrinsic SN Ia reddening-luminosity relation has a shallower
slope than the dust relation.  However, since the lightcurve
shape-luminosity relation is already factored out in this
method\citep{2006A&A...447...31A}, it means there must be a component
of intrinsic color that does not correlate with lightcurve
shape\citep{2007ApJ...664L..13C}.  
An alternative is that the dust along the line of sight to SNe Ia is
intrinsically different, or that scattering effects in the
circumstellar environment result in a different apparent reddening
law\citep{2008ApJ...686L.103G}.


One way to separate the intrinsic/dust reddening degeneracy is through
infrared observations, where the effects of dust are minimized.
Infrared observations have shown that many SNe seem to have a
sub-Milky Way value of $R_V$, sometimes as low as
$R_V=1.5$\citep{2008MNRAS.384..107E,2007AJ....133...58K}.  There are hints that lightly reddened SNe
may have an $R_V$ close to that in the Milky Way, while heavily
extincted SNe Ia have lower values (Fig. 3)\citep{2010AJ....139..120F}.
However, this is opposite to earlier findings from 
optical data\citep{2008A&A...487...19N}, and the small sample size makes it
difficult to draw reliable conclusions.

There are indications that supernovae with higher velocity ejecta have
lower inferred $R_V$ values\citep{2009ApJ...699L.139W}
(Fig.~\ref{redfig}).  Since the color is correlated with physical SN
features, this may indicate that sometimes an intrinsic color
difference in SNe is wrongly ascribed to dust reddening (though this
can be mitigated by making a color cut), or perhaps that there are
different progenitor scenarios, with different circumstellar dust
scattering properties, that produce SNe with different velocities.

Various authors have exploited the low sensitivity to dust in the IR
to make infrared Hubble diagrams.  This is doubly attractive, because
SNe Ia intrinsically show reduced scatter in the
IR\citep{2004ApJ...602L..81K} --- in the $H$ band SNe Ia have an rms
dispersion of only 0.15 mag {\it without} correction for lightcurve
shape or color\citep{2008ApJ...689..377W}.  This is theoretically
expected, in part because less luminous SNe are cooler and radiate a
larger fraction of their luminosity in the
IR\citep{2006ApJ...649..939K}.  While heroic attempts have been made
to produce high redshift IR Hubble diagrams of SNe
Ia\citep{2009ApJ...704.1036F}, the massive amount of observing time
required, and the fact that the restframe IR is redshifted to even
longer wavelengths, has made serious cosmological constraints from
this method elusive.



\subsection{Populations and evolution}
Even before SNe Ia were used for cosmology, it was known that the most
luminous SNe Ia (those with the broadest lightcurves), occur only in
late-type galaxies\citep{1996AJ....112.2391H}.  Likewise, subluminous
SNe Ia are preferentially found in galaxies with a significant old
population\citep{2001ApJ...554L.193H}.  SN lightcurve width and
luminosity have now been shown to correlate with host galaxy star
formation rate, galaxy mass, and
metallicity,\citep{2006ApJ...648..868S,2008ApJ...685..752G,
  2009ApJ...691..661H,2010MNRAS.406..782S}.

\begin{figure}
\includegraphics[width=7in]{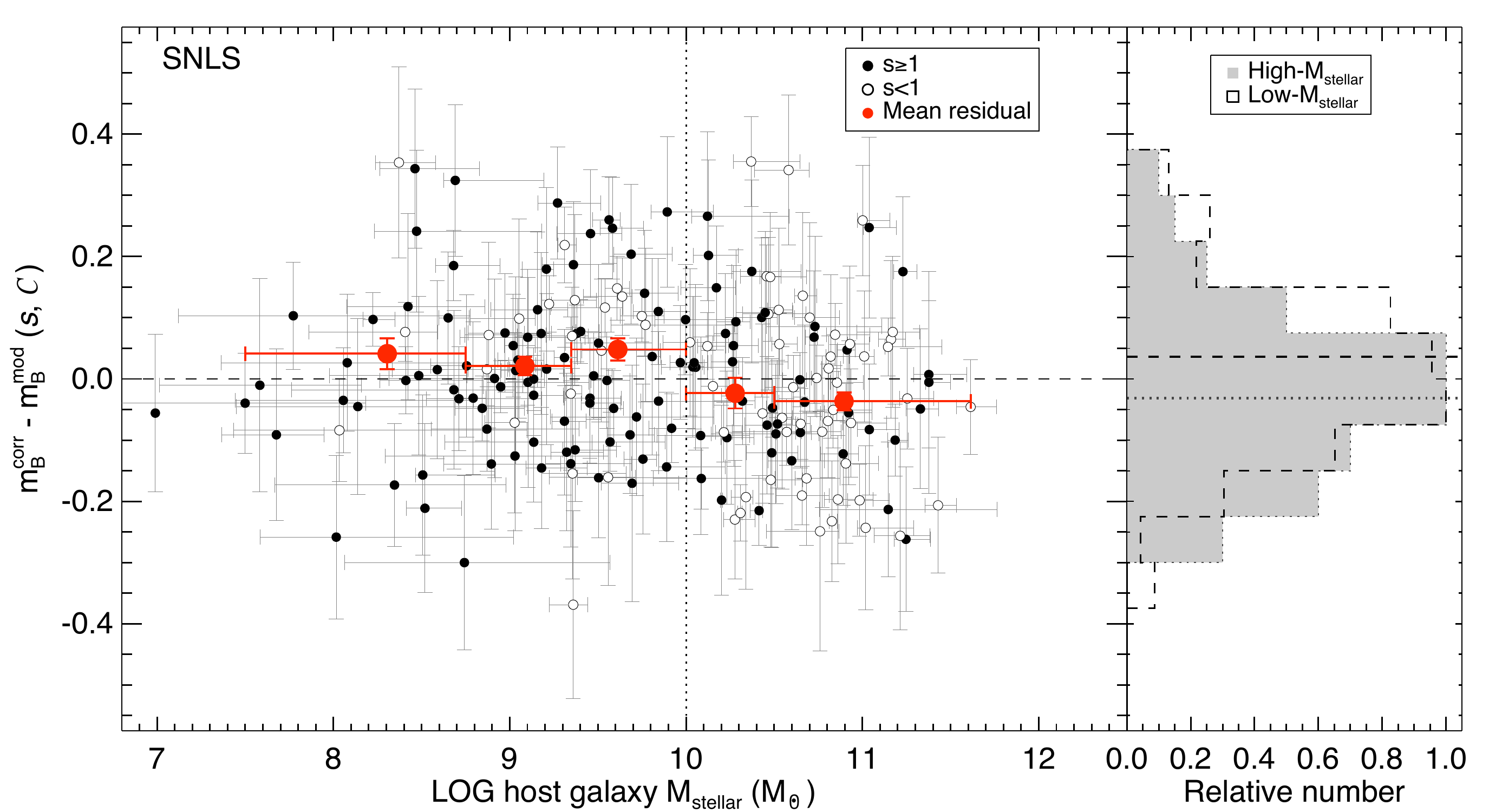} 
\caption{\textsf{{\bf \textsf{Residuals from the Hubble diagram for Supernova Legacy Survey Type Ia supernovae as a
    function of host galaxy mass {\it after} correction for lightcurve
    shape and color.}}  Figure from Sullivan et
  al.\citep{2010MNRAS.406..782S} Red points are averages.  Histograms
  on the right show the differences for high and low mass host
  galaxies, split at $M=10^{10}$ \Msun .  Horizontal lines show the
  average residual for each histogram.  SNe Ia in high and low mass
  hosts correct to an absolute magnitude different by $0.08\pm 0.02$
  mag ($4\sigma $).  Supernovae in low mass galaxies are on average
  brighter than those in high mass galaxies {\it before} correction,
  but are dimmer {\it after} $s$ and $c$ correction.  This
  galaxy-dependent residual can be corrected by taking host galaxy
  information into account.}}
\label{sullivan}
\vskip 0.05 in
\end{figure}

Since star formation increases by a factor of 10 up to redshift 1.5,
and bright, broad-lightcurve supernovae favor star forming hosts, it
is expected that the mix of supernovae will change with redshift.
Indeed it has been shown that supernovae at $z=1$ are intrinsically
$\sim 12\%$ more luminous than local SNe Ia
\citep{2007ApJ...667L..37H}.  The spectra of high-redshift supernovae
also show fewer intermediate mass elements, consistent with the idea
that they make more iron-peak elements to power their
luminosity\citep{2009ApJ...693L..76S}.  A changing mix of supernovae
with redshift is not necessarily problematic for cosmology if
lightcurve shape and color corrections allow all supernovae to be
corrected to the same absolute magnitude.  Unfortunately, they do not
\citep{2010MNRAS.406..782S,2010ApJ...722..566L} --- as is shown in
Fig.~\ref{sullivan}, supernovae in high and low mass galaxies each
correct to an absolute magnitude different by $0.08\pm0.02$ mag.  This
trend is also present (though weaker), if SNe are split by host star
formation rate, or metallicity.  If not corrected, this can lead to a
$\sim 0.04$ systematic error in $w$\citep{2010MNRAS.406..782S}.
Therefore, aside from color and lightcurve shape, a third correction,
one for host galaxy properties, must be applied to SNe Ia to avoid
systematic residuals with respect to the Hubble diagram.

The physical origins for the differences in supernova properties in
spiral and elliptical galaxies have been
elusive\citep{1996AJ....112.2391H,2001ApJ...554L.193H,2006ApJ...648..868S}
-- are they related to progenitor age, metallicity, or entirely
different progenitors?  The recently discovered trend that galaxy mass
may play the most significant role in determining SN
properties\citep{2009ApJ...691..661H,2010MNRAS.406..782S} would seem
to implicate metallicity, since higher mass galaxies retain more
metals in their deeper potential wells.  Still, the age-metallicity
degeneracy precludes firm conclusions on this evidence
alone\citep{2005ApJ...634..210G}.  Theoretically, there are reasons
that high metallicity progenitors should produce subluminous SNe
Ia\citep{2003ApJ...590L..83T}: increased $^{22}$Ne in high metallicity
white dwarfs provides more neutrons during the nucleosynthesis that
occurs during the explosion, and thus yields more stable $^{58}$Ni and
less of the radioactive lightcurve-powering $^{56}$Ni.  However, this
has only a $\sim 10\%$ effect on SN luminosity, perhaps indicating
that an as-yet unknown metallicity effect or progenitor age is the
more important in controlling \Ni\ yield and SN
luminosity\citep{2009ApJ...691..661H}.  

If progenitor age is the predominant effect controlling SN luminosity,
this has implications for SN progenitors.  In the single degenerate
(SD) scenario the time to explosion is controlled by the main sequence
lifetime of the secondary (mass transferring) star, since the mass
transfer timescale is thought to be negligible, only $\sim 10^7$
years.  Therefore, in this model, the delay time is a direct
indication of the secondary star's mass.  In the double degenerate
(DD) model, where two white dwarfs merge, gravitational wave radiation
ultimately brings the stars together, but they may have a head start
if they have drawn closer because at some time they have been orbiting
in the common envelope of one of the stars as it is evolving.  Age may
still play a role in determining luminosity, because at early times
the only white dwarfs will be massive ones derived from the more
massive stars.


\section{Explosion theory and observational constraints}\label{theorysec}
In what might be called the standard model for a SN Ia, a CO white
dwarf accretes matter until it compresses to the point that carbon is
ignited just before the Chandrasekhar limit (note that the common
misconception that the WD goes {\it over} the Chandrasekhar limit is
wrong -- this would lead to collapse to a neutron star).  The evidence
that the exploding star is a white dwarf is strong, albeit
circumstantial: neither hydrogen nor helium
is seen in the spectrum of a SN Ia\citep{2007ApJ...670.1275L}, SNe Ia
can happen long after star formation has ceased, the explosive process
may implicate degenerate matter, the energy obtained by the
thermonuclear burning of a white dwarf minus the binding energy
roughly matches the kinetic energy of SNe Ia, and simulations of the
process have been successful at reproducing SN Ia lightcurves and
spectra\citep{1984ApJ...286..644N,
  1991A&A...245..114K, 2000ARA&A..38..191H}.

The white dwarf may have a ``simmering'' phase
of order a thousand years following unstable carbon ignition, where
thermonuclear runaway is prevented by
convection\citep{2008ApJ...678.1158P}.  Ultimately, however, explosive
burning is ignited and the white dwarf is incinerated in seconds.  

\begin{figure}
\includegraphics[width=7in]{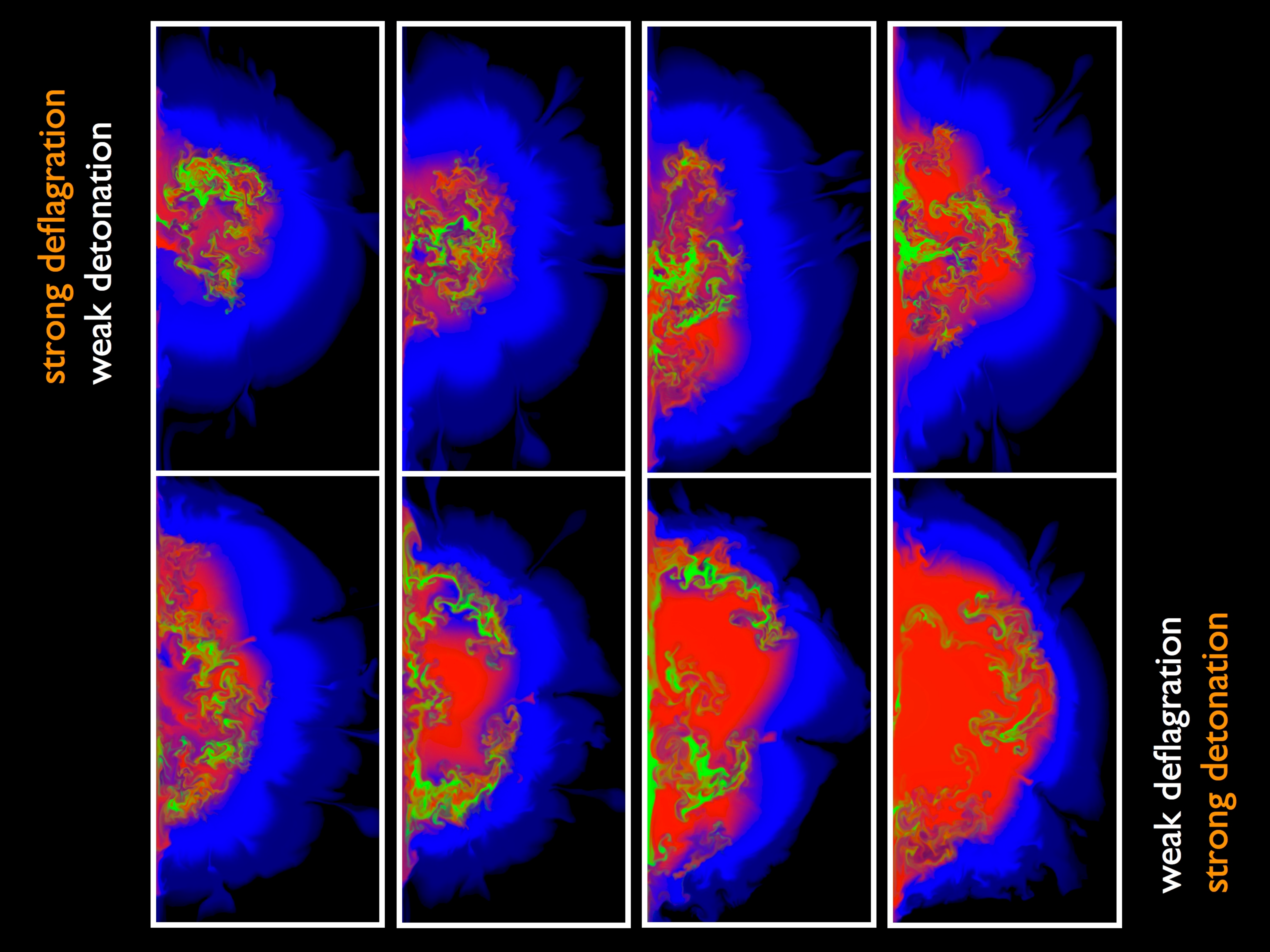} 
\caption{\textsf{{\bf \textsf{Two dimensional Type Ia supernova models showing the relative
    importance of deflagrations, detonations, and asymmetry.}}  Figure
  from Kasen et al.\citep{2009Natur.460..869K} Blue shows
  intermediate mass elements (Si, S, Ca), green is stable iron group
  elements, and red is \Ni .  An initial deflagration wave produces
  turbulent instabilities, but a later detonation wave burns much of
  the remaining fuel to \Ni .  In low density outer regions the
  detonation produces only intermediate mass elements.  Weak detonations produce the lower
  \Ni\ mass, less luminous SNe Ia.}}
\label{kasen}
\vskip 0.05 in
\end{figure}

If a white dwarf near the Chandrasekhar mass is detonated (i.e. the
burning occurs supersonically), then the white dwarf burns at such a
high density that the fusion products consist almost entirely of
iron-peak elements\citep{1969Ap&SS...5..180A}.  This does not match
the spectra or lightcurves of SNe Ia.  A deflagration (subsonic
burning), on the other hand, gives the SN time to pre-expand.  Burning
at a lower density can produce intermediate mass elements and
reproduce many of the observational features of SNe
Ia\citep{1984ApJ...286..644N}.  However, pure deflagrations fail to
produce the high velocity material (Fig.~\ref{fluxspec}) that seems to be
nearly ubiquitous in SNe Ia in the outer
layers\citep{2005ApJ...623L..37M}, and in two- and three dimensional
simulations they leave too much carbon and oxygen
unburned\citep{2005ApJ...623..337G}.  Therefore, the consensus is that
in a Chandrasekhar mass explosion, the flame must start out
subsonically, but at some point become
supersonic\citep{1991A&A...245..114K}.  Though the physics remains
poorly understood, models that start as a deflagration but impose a
transition to a detonation under certain conditions have been
successful at reproducing normal SN Ia lightcurves and
spectra, and even lightcurve
width-luminosity relations and metallicity
effects\citep{2009Natur.460..869K}.  Explosions dominated by
deflagration produce more intermediate mass elements (dimmer SNe Ia),
while those dominated by detonation produce brighter and more \Ni
-rich SNe Ia (Fig.~\ref{kasen})\citep{2009Natur.460..869K}.

It is possible to gain insight into the explosion physics with
spectropolarimetry or other techniques that reveal asphericity ---
recent work indicates that some of the dispersion in SN properties
results from broken symmetries\citep{2009Natur.460..869K}.  SN
asymmetry can be measured via spectropolarimetry, since asymmetric
electron scattering leads to polarization vectors that do not cancel.
Most normal SNe Ia are found to be spherically
symmetric\citep{2005ApJ...632..450L,2008ARA&A..46..433W}.  The first
convincing evidence for significant deviations from spherical symmetry
was seen in a subluminous SN Ia\citep{2001ApJ...556..302H}, which may
make sense if they are deflagration-dominated.  However, supernovae
with high velocity features often show even stronger
spectropolarimetric signatures of asymmetry, possibly due to clumpy
ejecta\citep{2003ApJ...593..788K,2005ApJ...632..450L}.  

Strangely, SNe Ia whose spectroscopic
features start off with high velocity and evolve rapidly, often show
nebular lines that appear redshifted, while SNe Ia with slower
velocity evolution show blueshifted nebular
lines\citep{2010Natur.466...82M}.  This probably indicates asymmetry
in the explosion, in qualitative agreement with models where a
deflagration burns off-center, and is followed by a detonation.


\section{The progenitor question\label{progensec}}
Even if there is agreement that the primary star in a SN Ia is a white
dwarf, the identity of the secondary star in the binary system is
uncertain.  There are three broad classes of models: (1) single
degenerate\citep{1973ApJ...186.1007W}, where the companion is a
main sequence or red giant star that loses mass via either Roche lobe
overflow, or a wind (symbiotic star), (2) double degenerate (DD),
where two white dwarfs merge and explode\citep{1984ApJS...54..335I,
  1984ApJ...277..355W}, and (3)
sub-Chandra\citep{1986ARA&A..24..205W}, where a layer of helium builds
up on the surface of a white dwarf below the Chandrasekhar mass until
it detonates.  


\subsection{Slow accretion}
In the single degenerate model, a white dwarf ignites carbon burning,
causing it to explode, by accreting matter from a nondegenerate
companion to near the Chandra mass.  Since CO white dwarfs are thought
to start no larger than $\sim 1-1.2$ solar masses, a significant
amount of accretion from a secondary star is required.  However
hydrogen must be accreted at a rate between $10^{-7}-10^{-8}$ \Msun\
yr$^{-1}$ to steadily burn to carbon or oxygen on the surface of the
white dwarf\citep{1982ApJ...253..798N}.  If it accretes faster, it
will form a red giant-like envelope.  Slower accretion is thought to
lead to the build up of matter that results in novae, and mass being
lost from the system.  The apparently required steady burning
generates supersoft X-rays, and as a result systems in this phase of
their evolution are observable as supersoft X-ray
sources\citep{1992A&A...262...97V}.  
However, the observed rates of supersoft sources in the Milky Way and
external galaxies can only account for $\sim 5\%$ of the SN Ia
rate\citep{2010ApJ...712..728D, 2010Natur.463..924G}, possibly
favoring the double degenerate model, although many of these systems
ought to pass through a supersoft phase as
well\citep{2010ApJ...719..474D}.


\subsection{White dwarf mergers}
In the double degenerate scenario, two white dwarfs merge to achieve
the Chandra mass\citep{1984ApJS...54..335I,1984ApJ...277..355W}.  The
less massive white dwarf is disrupted into a disk that will eventually
accrete onto the more massive white dwarf\citep{1990ApJ...348..647B}.
In this case, carbon may be ignited on the surface of the more massive
white dwarf during the accretion process, resulting in non-explosive
carbon burning, converting the star to an O-Ne white dwarf.  This
should ultimately undergo accretion induced collapse to a neutron star
rather than producing a SN Ia\citep{1985A&A...150L..21S}, though it
may be possible to avoid this fate under certain
conditions\citep{2007MNRAS.380..933Y}.  The simulated merger of two
equal mass 0.9 \Msun\ WDs has been shown to lead to a SN Ia, although
it was subluminous\citep{2010Natur.463...61P}, due to the low central
density of a 0.9 \Msun\ WD.  More massive equal-mass mergers may
produce normal, or even overluminous SNe Ia, but would be more rare.
Lower-mass mergers should be common, though it is not clear if they
lead to SNe Ia\citep{2010ApJ...722L.157V}.  The merger of two white
dwarfs is a very complicated three dimensional process (necessitating
approximations in current models), and very few simulations have been
completed.

The WD merger scenario has a natural explanation for greater SN luminosity
in young environments: younger, more massive stars produce more
massive white dwarfs.  Massive white dwarf mergers have more potential
fuel than less massive mergers.

\subsection{Sub-Chandrasekhar mass explosions}
While approaching the Chandra mass is a convenient method for triggering
carbon burning, there is no hard evidence that the Chandra mass is
required.  
In the classical sub-Chandra
models\citep{1986ARA&A..24..205W}, known as ``double detonation'' or
``edge-lit detonations,'' a layer of accreted helium ($\sim 0.2$\Msun
) is built up either by burning accreted hydrogen to helium or by
accretion from a helium rich donor.  When the pressure at the base of
the helium layer reaches a critical threshold, it detonates, driving a
shock into the core of the WD.  This causes a second detonation,
resulting in a flame propagating outward from the core (or near it),
destroying the WD.  Because the sub-Chandra white dwarf has a lower
density throughout, a simple detonation does not burn the entire star
to iron peak elements.  This model thus has the advantage that an ad
hoc deflagration to detonation transition is avoided.  However, the
significant outer helium layer is efficiently burned to $^{56}$Ni,
resulting in early time spectra that should be rich in \Ni , which
does not match observations\citep{1997ApJ...485..812N}.  If helium can
detonate with a smaller layer, which some studies
hint\citep{2010ApJ...715..767S}, then sub-Chandra simulations can
reproduce many of the observed properties of SNe
Ia\citep{2010ApJ...714L..52S}.  Some of the elegance of the model is
undercut by the required narrow range of WD masses: $\sim 0.9-1.1$
\Msun , in order to achieve the central densities necessary to produce
iron-peak elements.

\subsection{Constraints on progenitors from rates}
\begin{figure}
\includegraphics[width=7in]{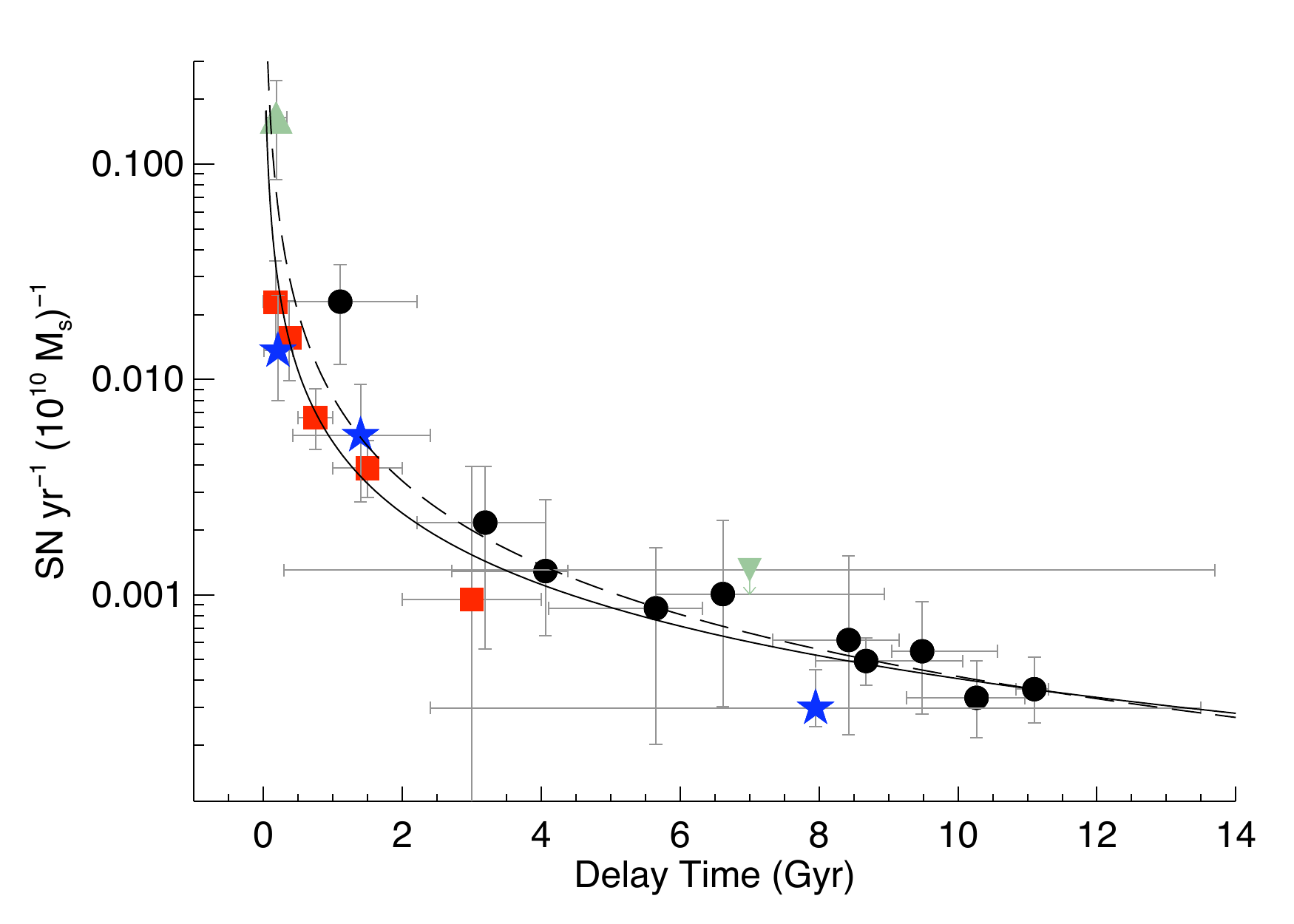} 
\caption{\textsf{{\bf \textsf{Delay time distribution for Type Ia supernovae.}}
    Adapted from Maoz, Sharon, and Gal-Yam\citep{2010ApJ...722.1879M}.
    All solid circles except the first are delay times derived from
    cluster SN rates, assuming formation at $z=3$.  The first solid
    circle results from a constraint on the observed iron-to-stellar
    mass ratio in clusters.  Blue stars are derived from SN rates and
    galaxy stellar populations in the Lick Observatory Supernova
    Survey\citep{2010arXiv1002.3056M}.  Green triangles (including an
    arrow denoting the 95\% confidence upper limit) are constraints
    from SN remnants in the Magellanic Clouds.  Red squares are based
    on SN Ia candidates in E/S0 galaxies at
    $z=0.4-1.2$\citep{2008PASJ...60.1327T}. Curves are power laws
    $t^{-1.1}$ (solid), and $t^{-1.3}$ (dashed), constrained to pass
    through the last point.  The prediction\citep{2005A&A...441.1055G}
    from DD mergers is $t^{-1}$}}
\label{maoz}
\vskip 0.05 in
\end{figure}

Different progenitor scenarios lead to different delay times between
the birth of the binary system and the explosion as a SN
Ia\citep{2005A&A...441.1055G}.  So there is hope of constraining the
progenitors by measuring the delay time distribution (DTD;
Fig.~\ref{maoz}).  One approach is to measure the lag between the
cosmic star formation rate and the SN Ia rate as a function of
redshift.  However, large uncertainties in each make this particularly
difficult, leading authors to different conclusions even
when using largely the same
data\citep{2010ApJ...713...32S,2008ApJ...673..981K}.  

Another approach is to use the relative SN rates in different types of
galaxies to constrain the DTD.  The fact that the SN Ia rate per unit
mass is higher by a factor of $\sim 20$ in late-type galaxies
indicates that
there is a significant population of short-lived thermonuclear SN
progenitors\citep{2005A&A...433..807M}.  While there has been much
argument about whether the SN Ia DTD is
bimodal\citep{2006MNRAS.370..773M}, a leap forward came with the
application of SED fitting techniques to SN host galaxies by modeling
their stellar populations\citep{2006ApJ...648..868S}.  This has
allowed more complicated DTDs to be determined: authors using likely
SNe Ia from the SXDS\citep{2008PASJ...60.1327T} (squares in
Fig.~\ref{maoz}), or those in clusters\citep{2010ApJ...722.1879M}
(filled circles), find a power law DTD proportional to $t^{-1}$.  It
is difficult to explain the power law DTD results using the single
degenerate scenario
alone\citep{2005A&A...441.1055G,2008PASJ...60.1327T}, though there is
at least one claim\citep{2008ApJ...679.1390H}.

Host galaxy spectroscopy contains even more information to constrain
progenitors.  New studies using different methods and different data sets
\citep{2010arXiv1002.3056M, 2010AJ....140..804B}, but
the same galaxy fitting code, have determined the relative
rates of supernovae in 0-0.42 Gy (prompt), 0.42 to 2.4 Gyr (medium),
and $>$2.4 Gyr (delayed) bins (stars Fig~\ref{maoz}).  They
find that ``prompt'' and ``delayed'' supernovae are required at
several sigma.  They also confirm earlier findings that bright,
broad-lightcurve supernovae favor a prompt population, while dim,
narrow-lightcurve SNe Ia favor a delayed population.  

Pinpointing the locations of SNe Ia relative to the stellar
populations in which they reside can give some information on
progenitors, even if it is imprecise due to the probability that the
SN has migrated from its birth site.
Comparing the distributions of SNe Ia to the blue light in galaxies
reveals that even ``prompt'' SNe Ia are significantly delayed, with
delay times 200-500 Myr\citep{2009ApJ...707...74R}.  Additional
constraints have been obtained by comparing SN Ia remnants in the
Magellanic Clouds (triangles in Fig.~\ref{maoz}) to their local
resolved stellar populations\citep{2010MNRAS.407.1314M}.  At a $>99\%$
confidence level, ``prompt'' SNe Ia that explode within 330 Myr of
star formation are required.

\subsection{Surviving or preceding material\label{stuff}}
If white dwarfs grow to become SNe Ia via hydrogen accretion, then it
ought to be possible to to see this hydrogen in the SN Ia in some
form.  However, despite decades of searching, no such smoking gun has
been found.  No SN Ia has ever been seen in the
radio\citep{2006ApJ...646..369P}, establishing an upper limit on
steady mass loss of material before the SN of $~\sim 3 \times 10^{-8}$
\Msun\ yr$^{-1}$.  This disfavors the symbiotic star hypothesis, where
the white dwarf accretes matter from the stellar wind of its
companion.   If, on the other hand, white dwarf is accreting
from Roche lobe overflow of a companion, the outer hydrogen layers of
the secondary ought to become stripped and entrained in the SN ejecta,
where it will show up at low velocities when the ejecta become
optically thin\citep{2000ApJS..128..615M}.  However, in a couple of
well observed SNe Ia, upper limits on the amount of hydrogen detected
are 0.01 \Msun , which would seem to rule out this
scenario\citep{2007ApJ...670.1275L}, at least for these SNe.  In a
few cases, apparent SNe Ia have been seen interacting with
pre-existing hydrogen, as in SN 2002ic\citep{2003Natur.424..651H},
although it is not clear these are really SNe
Ia\citep{2006ApJ...653L.129B}, and even if they are, these systems are
the exception rather than the rule.

The impact of the SN ejecta on any companion star would shock the
ejecta, leaving observational signatures along certain lines of sight
for several days, long after the ejecta has overtaken the
companion\citep{2010ApJ...708.1025K}.  This would lead to distortions
in the early-time lightcurves of SNe Ia around 10\% of the time,
especially in the UV to $B$ regions of the spectrum.  If the companion
star is a red giant, particularly large distortions ought to be seen --
the fact that they never have been would seem to be problematic for
this scenario\citep{2010ApJ...722.1691H}.

Variable sodium lines have been seen in high resolution spectra of
some SNe Ia, and their variability, assumed to be from ionization of
circumstellar clouds by the SN, has been taken to be evidence of
significant circumstellar material.  This ostensibly favors the single
degenerate scenario\citep{2007Sci...317..924P}, although maybe only in
rare cases --- variable sodium was only seen in 6\% of cases in a
sample of 31 observations, and then only in the most reddened SNe
Ia\citep{2009ApJ...693..207B}.  Not all variable sodium SNe Ia are
reddened, however, nor do all reddened supernovae show variable
sodium\citep{2009ApJ...702.1157S}.  There have also been doubts that
this material is truly circumstellar\citep{2008AstL...34..389C}.


SN remnants provide more clues to the origin of SNe Ia.  They may
reveal the progenitor's metallicity\citep{2008ApJ...680.1149B}, for
example.  But even more exciting is the prospect that they may house a
surviving progenitor star.  Tycho's remnant (SN 1572), long suspected
of being a SN Ia, has been definitively revealed to be so by modern
spectroscopic observations of the SN light, delayed in transit by
reflections off of distant interstellar
clouds\citep{2008Natur.456..617K}.  In the single degenerate scenario,
the secondary star should survive the SN explosion, but may be
stripped of its outer layers, and heated, such that it takes of order
1000 years to return to thermal equilibrium\citep{2000ApJS..128..615M}.
One star, Tycho G, has been claimed as evidence for a surviving
progenitor\citep{2004Natur.431.1069R}, though
there are many doubts about this hypothesis.  It is too
displaced from the center of the remnant, and its proper motion shows
it did not come from that direction.  Also, it was not apparently out
of thermal equilibrium, and does not have the expected rotational
characteristics\citep{2009ApJ...701.1665K}.

\subsection{Super-Chandra}
Beginning with SNLS-03D3bb\citep{2006Natur.443..308H}, aka SN 2003fg,
several supernovae have shown luminosities indicating a Nickel yield
alone of 1.3 -- 1.8 \Msun \citep{2010ApJ...713.1073S,2011MNRAS.tmp...61T}, possibly indicating a super-Chandrasekhar mass
progenitor.  Such energetic explosions ought to have more kinetic
energy, and thus higher velocity ejecta, but this class of SNe
counterintuitively have much slower than average ejecta.  This may
result from the increased binding energy of a super-Chandra
WD\citep{2006Natur.443..308H}, or from the ejecta plowing into a dense
shell of circumstellar material resulting\citep{2010ApJ...713.1073S}.
These SNe also show rare lines of CII near maximum light, possibly
indicating less complete burning.  They may also favor a young stellar
population.  This class of SNe Ia would seem to implicate the merger 
of two white dwarfs, but no simulation has yet reproduced one.

\section{Final Thoughts}\label{finalsec}
SN Ia are mature probes of cosmology: systematic uncertainties now dominate
statistical uncertainties up to $z=1$, and the systematic uncertainties are well
characterized.  Still, there is plenty of room for improvement in the
use of SNe Ia as standard candles.  

It is possible that no single model can describe all SNe Ia.  Perhaps
some SNe Ia are sub-Chandra, some are the result of mergers, and some
are caused by Roche Lobe overflow.  The explosion of a white dwarf
may be similar no matter how it is triggered.  If it is not required
that a model explain all SNe, this opens the door to scenarios that do
not have high expected event rates.

In the study of SNe Ia, we have left the ``serendipity era,'' when we
could only discover what nature happened to tell us through chance
encounters with nearby supernovae, and we are entering the ``database
era,'' where we can proactively ask and answer questions about
supernovae by comparing large subsamples.  The legacy yields of the
multi-year second-generation SN surveys 
are just becoming apparent.  Meanwhile,
some third-generation surveys like the Palomar Transient Factory,
Pan-STARRS, the Dark Energy Survey, and Skymapper, have already
started to produce results, and promise thousands more SNe Ia in the
next few years.  Finally, fourth-generation surveys, the Large
Synoptic Survey Telescope (LSST), and possible space missions promise
unimaginable riches on the horizon.  New techniques like comparing
large subsamples, using UV and IR data, state of the art calibration,
using host galaxy observations, and 3d modeling, have started to yield
results, but are still in their infancy.  The next decade holds real
promise of making serious progress understanding nearly every aspect
of SNe Ia, from their explosion physics, to their progenitors, to
their use as standard candles.  And with this knowledge may come the
key to unlocking the darkest secrets of dark energy.

\bibliography{astro}

\begin{center}
\begin{table}
\begin{tabular}{lllll}
\hline
Systematic & SNLS3\citep{2011ApJS..192....1C,2011arXiv1104.1444S} & CfA\citep{2009ApJ...700.1097H}/ESSENCE\citep{2007ApJ...666..694W} & SDSS-II\citep{2009ApJS..185...32K}&SCP\citep{2010ApJ...716..712A}\\
\hline
Best fit $w$ (assuming flatness)&-1.061& -0.987&-0.96  &-0.997 \\
Statistical error               &\nodata& 0.067 & 0.06  & 0.052 \\
Total stat+systematic error     &0.069& 0.13  & 0.13  & 0.08  \\
\hline
Systematic error breakdown\\
Flux reference                 &0.053  &0.02   & 0.02  &0.042  \\
Experiment zero points         &0.01   &0.04   & 0.030 &0.037  \\
Low-z photometry               &0.02   &0.005  &\nodata&\nodata\\
Landolt bandpasses             &0.01   &\nodata& 0.008 &\nodata\\
Local flows                    &0.014  &\nodata& 0.03  &\nodata\\
Experiment bandpasses          &0.01   &\nodata& 0.016 &\nodata\\
Malmquist bias model           &0.01   &0.02   &\nodata &0.026 \\
Dust/Color-luminosity ($\beta$)&0.02   &0.08   &0.013  &0.026  \\
SN Ia Evolution                &\nodata&0.02   &\nodata&\nodata\\
Restframe U band               &\nodata&\nodata&0.104  &0.010  \\
Contamination                  &\nodata&\nodata&\nodata&0.021  \\
Galactic Extinction            &\nodata&\nodata&0.022  &0.012  \\
\hline
\end{tabular}
\caption{{\bf \textsf{Best-fit values of $\langle w\rangle$ and error estimates.}}  For the CfA3/ESSENCE column, $w$ is from Hicken et al. 2009\citep{2009ApJ...700.1097H}, though uncertainties are from ESSENCE\citep{2007ApJ...666..694W}, as they are stated to be similar but are 
  not separately tabulated.  The SDSS numbers\citep{2009ApJS..185...32K} 
  are for their SALT2 fit.  Errors for each survey use their largest sample.  
 \label{tabsys}
}
\end{table}
\end{center}

\begin{figure}
\includegraphics[width=7in]{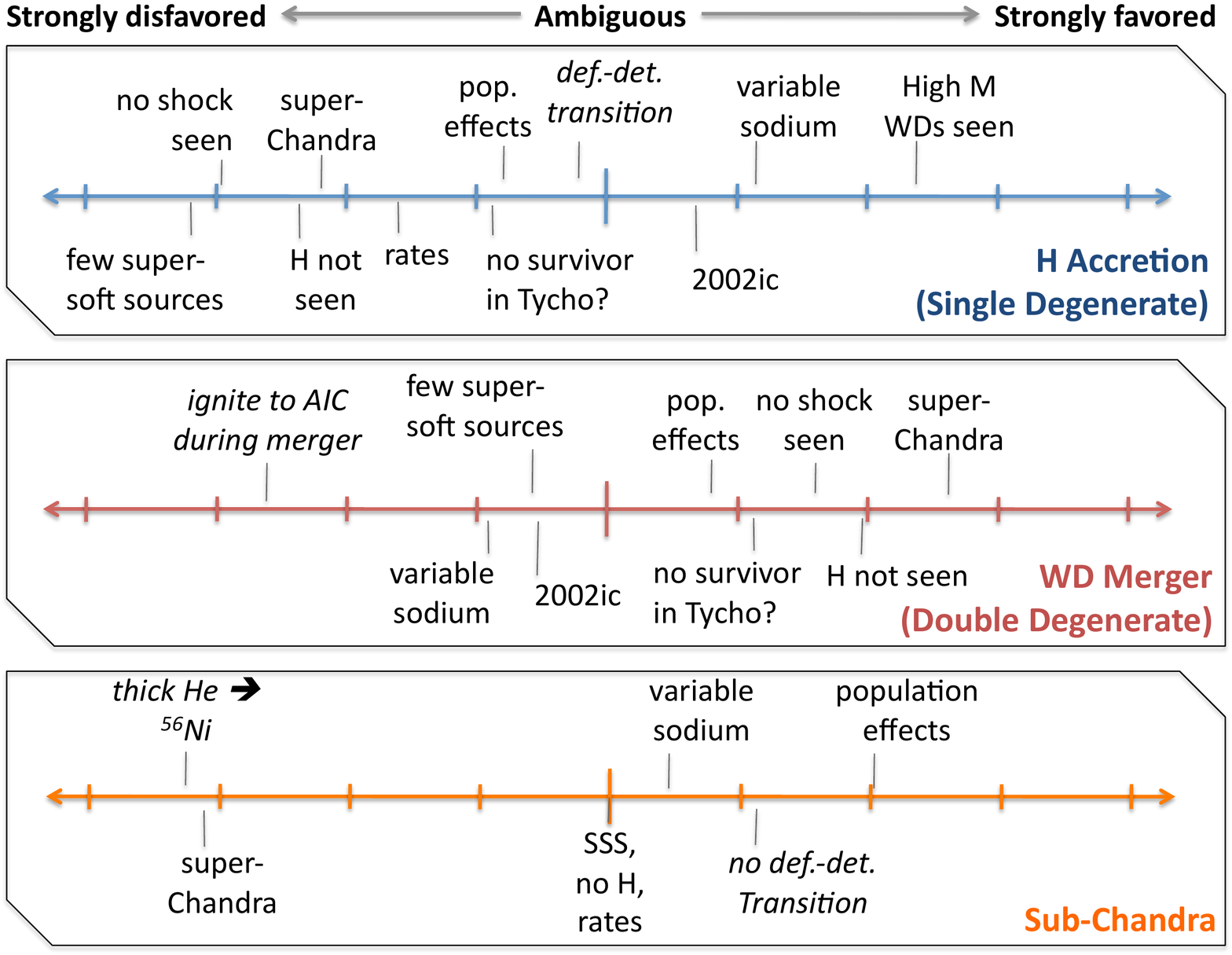} 
\caption{\textsf{{\bf \textsf{Schematic illustration of the arguments for and against a
  given progenitor scenario being the predominant mechanism for SNe Ia.}}  Arguments on the left strongly disfavor a certain scenario,
  while arguments on the right strongly favor that model.  Note
  that an argument against a given scenario need not be an equally
  strong argument for another scenario (i.e. the diagram need not be
  symmetric).  Though all evidence is subject to theoretical
  interpretation, italics indicate exclusively theoretical arguments.
  Relative rankings reflect an effort to distill community consensus,
  but are subjective.  Arguments and references are described in the text.  ``Population effects'' refers to the fact that more luminous SNe Ia are found in younger environments.  ``High M WDs seen'' refers to the fact that CO WDs $> 1.2$ \Msun\ exist in nature, thought to only be possible by slow accretion.
}}
\label{evidence}
\vskip 0.05 in
\end{figure}



\begin{addendum}
 \item DAH acknowledges support from LCOGT.
 \item[Competing Interests] The authors declare that they have no
competing financial interests.
\item[Correspondence] Correspondence should be addressed to
  DAH~(email: ahowell@lcogt.net).
\end{addendum}


\end{document}